\RequirePackage{fix-cm}
\documentclass[smallextended]{svjour3}       
\smartqed  
\usepackage{graphicx}
\usepackage{multirow}
\usepackage{footnote}
\usepackage{tablefootnote}
\newcommand{\quotes}[1]{``#1''}

\begin{document}

\title{Analysis of Search Stratagem Utilisation}




\author{Ameni Kacem \and Philipp Mayr}


\institute{Ameni Kacem  \\
              \email{ameni.sahraoui@gesis.org}           \at
              GESIS -- Leibniz Institute for the Social Sciences \\
              Cologne, Germany\\
              https://orcid.org/0000-0001-8728-9406\\
           \and
          \\ Philipp Mayr \\
           \email{philipp.mayr@gesis.org}  \at
              GESIS -- Leibniz Institute for the Social Sciences \\
              Cologne, Germany\\
              https://orcid.org/0000-0002-6656-1658\\
}

\date{Received: date / Accepted: date}

\maketitle

\begin{abstract} 
In Interactive Information Retrieval, researchers consider the user behaviour towards systems and search tasks in order to adapt search results and to improve the search experience of users. Analysing the users' past interactions with the system is one typical approach. In this paper, we analyse the user behaviour in retrieval sessions towards Marcia Bates' search stratagems such as \quotes{Footnote Chasing}, \quotes{Citation Searching}, \quotes{Keyword Searching}, \quotes{Author Searching} and \quotes{Journal Run} in a real-life academic search engine. In fact, search stratagems represent high-level search behaviour as the users go beyond simple execution of queries and investigate more of the system functionalities.  
We performed analyses of these five search stratagems using two datasets extracted from the social sciences search engine sowiport. A specific focus was the detection of the search phase and frequency of the usage of these stratagems. In addition, we explored the impact of these stratagems on the whole search process performance. 
We addressed mainly the usage patterns' observation of the stratagems, their impact on the conduct of retrieval sessions and explore whether they are used similarly in both datasets. 
From the observation and metrics proposed, we can conclude that the utilisation of search stratagems in real retrieval sessions leads to an improvement of the precision in terms of positive interactions. For both datasets (SUSS 14-15 and SUSS 16-17), the user behaviour was similar as all stratagems appear most frequently in the middle of a session. However, the difference is that \quotes{Footnote Chasing}, \quotes{Citation Searching} and \quotes{Journal Run} appear mostly at the end of a session while Keyword and Author Searching appear typically at the beginning. Thus, we can conclude from the log analysis that the improvement of search functionalities including personalisation and/or recommendation could be achieved by considering references, citations, and journals in the ranking process.
 


\end{abstract}

\section{Introduction}
\label{sec:intro}
Interactive Information Retrieval (IIR) refers to a research discipline that studies the interaction between the user and the search system. In fact, researchers have moved from considering only the current query and result set to focus more on the user's past interactions and the analysis of whole retrieval sessions \cite{Dumais:2016,Ruthven:2008,Kelly:2009}. A group of researchers is formulating that there is \quotes{a need for whole-session evaluation} in IIR\footnote{See \quotes{Whole-Session Evaluation of Interactive Information Retrieval Systems} seminar, http://shonan.nii.ac.jp/shonan/seminar020/.}. This requirement is very much in line with the research presented in this paper.
Following the \quotes{whole-session evaluation} desideratum, research approaches aim to understand the user searching behaviour in order to improve the ranking of results after submitting a query and enhance the user experience within an IR system.

In Digital Libraries (DLs), IIR researchers study concepts such as search strategies \cite{Marchionini1992,Belkin1993,Kriewel/Fuhr:10,Carevic:2016}, search term suggestions \cite{Belkin2001,shiri2002,Hienert:2016,Mayr2016BIRNDL}, user modelling \cite{Tran:2017,Dungs:2017}, communities' detection \cite{Akbar:2012}, personalisation of search results \cite{Rohini2005,Liu2012}, recommendation's impact \cite{Hienert:2016}, user's information needs change \cite{Xie2002} and many more topics. 
In addition, many interactive IR models have been proposed in the literature (e.g. \cite{Ellis:1989}) that describe the user behaviour by different steps (stages) of information seeking and interacting with an IR system. 
Exploratory search in DLs and academic search engines \cite{Carevic2017} is a rewarding research environment for interactive IR researchers because evolving searches with complex search tasks can be observed much easier compared to web search where searchers often jump into different websites. In DLs, users typically stay in the system and work with the variety of facilities it offers. This is due to the fact that state-of-the-art DLs offer dozens of possibilities to navigate and interact with the search system \cite{Hienert2015,Fuhr2007}.

Similarly, in the academic search engine sowiport \cite{Hienert2015}, we aim at understanding the user behaviour in order to support him/her during the search session. In fact, DL users behave differently when interacting with the system as underlined by Bates \cite{Bates,Bates1993LQ,Bates1993_jasis} who highlighted different concepts such as moves, tactics, stratagems, and strategies. 

The goal of this paper\footnote{This paper is an extended version of the paper \quotes{Analysis of Footnote Chasing and Citation Searching in an Academic Search Engine} presented at the Bibliometric-enhanced Information Retrieval and Natural Language Processing for Digital Libraries (BIRNDL) workshop at SIGIR 2017 \cite{Kacem2017}.} is to explore the specific stratagems \quotes{Footnote Chasing}, \quotes{Citation Searching}, \quotes{Keyword Searching}, \quotes{Author Searching} and \quotes{Journal Run} which are often utilised as exploratory search functionalities in DLs \cite{Carevic:2016,Carevic2017}. The relationship between Information Retrieval and Scientometrics has been discussed in a recent special issue in Scientometrics \cite{mayr2015}. In the current paper we follow the argument of \cite{mayr2015} that \quotes{at the root of any scientometric or bibliometric study} there is a comprehensive information retrieval task for the bibliometrician (researcher in bibliometrics) to gather the necessary documents to perform proper scientific research. Understanding and utilising stratagems, as proposed in this paper, can have a strong influence on the effectiveness of the retrieval task, and of course the precision and recall a bibliometrician can achieve. Concrete bibliometric-enhanced retrieval services have been proposed in a previous work \cite{mutschke2011}.   

\begin{figure}
\begin{center}
	\includegraphics[width=1.0\textwidth]{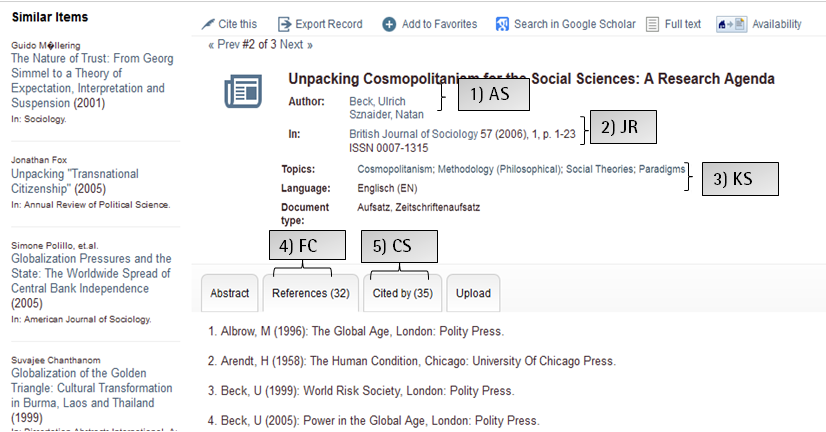}
	\caption{Search stratagems implemented in sowiport: Author Searching (1=AS), Journal Run (2=JR), Keyword Searching (3=KS), Footnote Chasing (4=FC) and Citation Searching (5=CS) as seen from a typical seed document.}
	\label{fig:searchstrat}
\end{center}
\end{figure}

According to Bates \cite{Bates}, \quotes{Footnote Chasing} is defined as checking the cited references and related material of a work backward in time after checking a document. \quotes{Citation Searching} refers to a forward chaining of works citing the seed document through a citation index. \quotes{Keyword Searching} or \quotes{Area Scanning} consists of looking up the indexing terms representing research topics after finding an area of interest in a classification system. The fourth stratagem \quotes{Author Searching} is defined as looking for specific author names to investigate more written material from a concrete author. Regarding the stratagem \quotes{Journal Run}, it was defined as browsing documents within a specific journal after its identification as relevant to the user's topic of interest.
These stratagems are important search features which are built in state-of-the-art academic search engines like Google Scholar or ACM Digital Library and support the natural search behaviour of a majority of academic searchers. The goal of this paper is to analyse the utilisation of the most popular search stratagems in an academic search engine in the social sciences, the sowiport search engine (see Figure \ref{fig:searchstrat} with an overview of all implemented stratagems of this study).

Stratagems in general are not always supported by DLs because most of the search functions available in academic search engines remain on the \quotes{moves} or \quotes{tactics} level (described in \cite{Bates}) or metadata like cited references are completely missing in the system.
However, we investigate in the following if the use of stratagems can enhance the search experience of the user, and which stratagems could be integrated in an academic search engine interface as well as in the re-ranking process of scientific papers (e.g. for personalisation or contextualisation).

In particular, we address the following research questions:\\ 

\textbf{RQ 1: Which usage patterns can be observed from specific search stratagems or moves?}
In this study, we analyse the usage patterns of specific search stratagems in real retrieval sessions in terms of frequency of their use and the stage at which they appear. More precisely, we examine the user behaviour towards five types of stratagems: \quotes{Footnote Chasing}, \quotes{Citation Search}, \quotes{Keyword Search}, \quotes{Author Search} and \quotes{Journal Run}. 

\textbf{RQ 2: How successful are the retrieval sessions which contain specific stratagems?}
The use of a stratagem can impact the session conduct in different ways. We examine the interactions of the users in the DL sowiport in order to measure the usefulness and the precision of sessions containing such stratagems. We determine the session success based on the presence of positive actions proposed recently by Hienert and Mutschke \cite{Hienert:2016}. In particular, we measure the percentage of positive actions, considered as positive implicit relevance feedback or indicators, before and after the stratagems' occurrence in a retrieval session. 

\textbf{RQ 3: How significant are the changes in the usage of stratagems using the same retrieval system in two different periods?}
We analysed the usage of the same search stratagems using two different datasets extracted both over a period of one year from the DL sowiport. We propose to study the frequency and stage of stratagems as well as the impact of their usage and analyse the differences between the user behaviours for both of the datasets.\newline

The remainder of this paper is organized as follows. In section~\ref{sec:related}, we present an overview of basic research conducted mainly by researchers in the field of DLs. In section~\ref{sec:Methodology}, we introduce the datasets and methodology of the study. We analyse the user behaviour towards specific stratagems and how using them affects the quality of the whole-session search in section~\ref{sec:results}. Finally, we summarise our findings and present some perspectives for future work.

\section{Related Work}  
\label{sec:related}
In this section, we investigate related work regarding log analysis and stratagem analysis.  
\subsection{Transaction Log Analysis}
Many research works have been proposed in the literature that analyse the information seeking process of users in search engines or any information system. Transaction logs are typically used to identify usage trends but also to find weaknesses in a system like usability bugs. A transaction log is defined as a file of the transactions (communications) between a system and its users \cite{Rice:1983}. 
Transaction log analysis (TLA) is the process of knowledge extraction and relevant information extraction from available data (logs). These analyses allow also understanding the interaction between the user and such system and recognise search patterns. Jansen et al. \cite{Jansen2006} provide also the possible applications of TLA, its limitations and its process a) \textit{Data Collection} including possible features such as user identification, date, search URL et cetera, b) \textit{Data Preparation} which consists in importing data to a relational database or other analysis mechanism, and includes data cleaning, parsing as well as normalising, and c) \textit{Data Analysis} which contains three levels of analysis in their work, namely, term level analysis, query level analysis, and session level analysis.
Agosti et al. \cite{Agosti:2012} identify two main different lines: Web search engines log analysis and Digital Library Systems log analysis. According to the authors, in the latter category, the researchers consider a) lab studies where they can easily observe the user's tasks, b) instrumented panels where the users accept to take part in a periodic study using for instance a browser application, and c) the aggregation of query logs where records contain not only the query but also other actions performed by the user.

\subsection{User Behaviour Analysis}
\label{sec:user behaviour analysis}
Bates \cite{Bates} has specified different types of user behaviour toward search system, among them we cite: \textit{moves}, \textit{tactics}, \textit{stratagems} and \textit{strategies}. A move refers to a basic action performed by the user. A tactic resides in using additional moves to go with the search. As for stratagems, they indicate complex and multiple moves/tactics having knowledge of a particular search domain. A strategy is a combination of moves, tactics and stratagems as a plan to pursue during the search session.
Many approaches studied the user behaviour towards tactics, moves or stratagems. For instance, Schneider and Borlund \cite{Schneider:2004} studied the effectiveness of using stratagems in constructing and maintaining thesauri vocabulary and structure.
Mahoui and Cunningham \cite{mahoui:2001} specified the importance of understanding the information of DL users in creating useful and stable search systems. 
They analysed transaction logs to study usage patterns of CiteSeer in terms of query and search patterns.
Xie \cite{Xie2002} analysed the user's search behaviour and their relationships with their information needs by specifying a hierarchical level of the users' goals.
Shute and Smith \cite{SHUTE1993} identified 13 knowledge-based tactics arranged into three categories: broaden topic scope, narrow topic scope and change topic scope.
Carevic and Mayr \cite{Carevic:2015} proposed bibliometric-enhanced search facilities such as \quotes{Journal Run} or \quotes{Citation Search} and their possible integration in DLs. In their position paper, they argue that bibliometric-enhanced stratagems can facilitate domain specific search activities by applying bibliometric measures for re-ranking and/or rearranging DL-entities like documents, journals or authors. They propose different types of stratagem implementations like \quotes{extended journal run} or \quotes{context-preserving journal run} or extended versions of \quotes{Citation Search}.

In \cite{Carevic2017}, the authors study exploratory search tasks that require users to investigate obtained results in DLs. They focused on early stages: \quotes{starting}, \quotes{chaining}, and \quotes{browsing}. In addition to the eye tracking analysis, they performed a user study based on a given task in a DL with two groups of users (students and postdoctoral researchers). Similarly, Xie et al. \cite{Xie2017} studied different types of search tactics. They compared user-dominated, system-dominated, and balanced tactics using the search process and they defined \quotes{creating}, \quotes{exploring}, \quotes{evaluating}, etc. In addition, a categorization of the user behaviour has been proposed including perceptional and behavioural measures \cite{Zhuang}. Some approaches focus on the evaluation of IIR systems such as the work of Borlund \cite{Borlund:2016A,Borlund:2016B} where the author proposed to consider types of information needs and introduced a meta-evaluation through the study of the simulated work task situations based on citation analysis through Web of Science\footnote{http://apps.webofknowledge.com/} and ACM Digital Library\footnote{https://dl.acm.org/}. 

In this paper, we are interested in studying five types of stratagems defined in \cite{Bates,Bates2}, namely: \quotes{Footnote Chasing}, \quotes{Citation Searching}, \quotes{Keyword Searching}, \quotes{Author Searching} and \quotes{Journal Run}. In fact, \quotes{Footnote Chasing} and \quotes{Citation Searching} are popular stratagems that refer to the study of documents and their bibliographic references and citations. \quotes{Keyword Searching} refers to the terms indexed in each document description. Other users search for author or editor names in the query which is referred as \quotes{Author Searching}. \quotes{Journal Run} refers to the examination of papers present in a specific source (in this case the source is a journal).

\section{Methodology} 
\label{sec:Methodology}
In this section, we first provide details about the dataset that we used for our analysis. Then, we describe the approach used to answer the research questions raised in Section~\ref{sec:intro}. 

\subsection{Datasets} 
\label{sec:dataset}
Sowiport is a DL for the Social Sciences that contains more than nine million records, full texts and research projects collected from twenty-two different databases whose content is in English and German \cite{Hienert2015}. For a part of the collections, namely the ProQuest databases \quotes{Sociological Abstracts}, \quotes{Social Services Abstracts}, \quotes{Applied Social Sciences Index and Abstracts}, \quotes{Worldwide Political Science Abstracts} and \quotes{Physical Education Index}, sowiport provides references and builds a citation index over its collections. These references and citations are part of the analysis in the following.

The \textit{Sowiport User Search Sessions Dataset (SUSS)}\footnote{http://dx.doi.org/10.7802/1380 and \cite{Mayr2016}} contains individual search sessions extracted from the transaction logs of sowiport. The first dataset was collected over a period of one year (between 2nd April 2014 and 2nd April 2015)\footnote{A detailed description of the dataset can be found in \cite{Mayr2017}.} denoted as \textit{SUSS 14-15}. The second dataset was collected from September 2016 to May 2017 denoted in this paper as \textit{SUSS 16-17}\footnote{The dataset can be downloaded at https://git.gesis.org/amur/SUSS-16-17} and is intended to generalise and validate the results from \textit{SUSS 14-15}.

The Web server log files and specific JavaScript-based logging techniques were used to capture the user behaviour within the system. The log was heavily filtered to exclude transactions performed by robots. All transaction activities are mapped to a list of 58 different user actions which cover all types of activities and pages that can be carried out/visited within the system (e.g. typing a query, visiting a document, selecting a facet, exporting a document, etc.). For each action, a session id, the date stamp and additional information (e.g. query terms, document ids, and result lists) are stored. Based on the session id and date stamp, the step in which an action is conducted and the length of the action is included in the dataset as well. The session id is assigned via browser cookies and allows tracking the user behaviour over multiple searches. We present in Table~\ref{tab:datasets} both datasets in terms of number of logs entries, individual searches, and registered users.

\begin{table}
\centering
\caption{Characteristics of both datasets SUSS 14-15 and SUSS 16-17}
\label{tab:datasets}
\begin{tabular}{lll}
\hline\noalign{\smallskip}
 & SUSS 14-15 & SUSS 16-17 \\
\noalign{\smallskip}\hline\noalign{\smallskip}
Number of Log Entries & 7,982,427 & 3,376,997  \\
Number of Individual Searches   & 484,449  &  208,556 \\
Number of Registered Users & 1,509 & 709  \\
Sessions Performed by Registered Users &  3,372 & 1,317 \\
Average Session Length (s) & 2,664 & 2,911  \\
Average Number of Actions & 16 & 16   \\
\noalign{\smallskip}\hline
\end{tabular}
\end{table}

\subsection{Description of Actions in the Session Log}
Searching sowiport can be performed through an \textit{All fields} search box (default search without specification), or through specifying one or more field(s): title, person, institution, number, keyword or year.
The users' main actions are described in Table~\ref{tab:actions}. In fact, we grouped the main actions into two categories: \quotes{Query}-related and \quotes{Document}-related actions. Another categorisation of actions was proposed in \cite{Hienert:2016} by specifying search interactions and successive positive actions. 

\begin{table}
\centering
\caption{Query- and document-related actions performed by users in sowiport}
\label{tab:actions}
\begin{tabular}{llp{7cm}}
\hline\noalign{\smallskip}
Category & Action & Description \\
\noalign{\smallskip}\hline\noalign{\smallskip}
 \multirow{8}{*}{Query} & query\_form & Formulating a query \\
    & search & A search result list for any kind of search \\
    & search\_advanced  &  A search with the advanced settings that can limit the search fields, information type, provider/database, language: or time (year, recent only) \\
    & search\_keyword & A search for a keyword  \\
    & search\_thesaurus & Usage of the thesaurus system\\
    & search\_institution & A search for an institution \\
    & search\_person & A search for a specific person (author/editor) \\
\noalign{\smallskip}\hline\noalign{\smallskip}
\multirow{8}{*} {Document} & view\_record & Displaying a record in the result list after clicking on it\\
    & view\_citation & View the document's citation(s) \\
    & view\_references & View the document's references \\
    & view\_description & View the document's abstract \\
    & export\_bib & Export the document through different formats \\
    & export\_cite & Export the document's citations list \\
    & export\_mail & Send the document via email \\
    & to\_favorites & Save the document to the favorite list \\
\noalign{\smallskip}\hline
\end{tabular}
\end{table}

Main user actions as described before can be categorised into actions regarding either search queries or documents. These actions are used in different scales in the datasets. Query-related actions represent 29.84\% in the SUSS 14-15 dataset and 29.63 \% in the SUSS 16-17 dataset, while document-related actions represent 35.79\% of the total amount of actions for the SUSS 14-15 dataset and 16.22\% for SUSS 16-17. The rest of actions contains navigational interactions such as logging in the system, managing favorites, and accessing the system pages.
 
In Table~\ref{tab:SessionSample}, we show a specific session, the user's ID and the action label and length in seconds. 
In this session, the user with ID \textit{41821} started with logging into the system and then submitted a query describing his/her information need (\textit{query\_form}) after performing some navigational actions. After getting the result list, labelled as \textit{resultlistids}, the user performed additional searches (\textit{searchterm\_2}), and displayed some results' content (\textit{view\_record}). Finally, he/she checked the external availability of a result (\textit{goto\_google\_scholar}). We notice that the user spent more than 40\% of the time reading the documents' content.

\begin{table}[!h]
\centering
\caption{Sample of a session search for a specific user}
\label{tab:SessionSample}
\begin{tabular}{lp{3cm}p{3.5cm}l}
\hline\noalign{\smallskip}
User ID & Date & Action label & Action length (s)\\
\noalign{\smallskip}\hline\noalign{\smallskip}
\multirow{9}{*} {41821} & 2014-10-28 16:08:46 & goto\_login   & 1   \\
& 2014-10-28 16:08:47 & goto\_favorites       & 21  \\
& 2014-10-28 16:09:08 & goto\_home            & 2   \\
& 2014-10-28 16:09:13 & query\_form           & 22  \\
& 2014-10-28 16:09:35 & search                & 10  \\
 & 2014-10-28 16:09:35 & searchterm\_2         & 10  \\
& 2014-10-28 16:09:35 & resultlistids         & 10  \\
 & 2014-10-28 16:09:45 & view\_record          & 31  \\
& 2014-10-28 16:09:45 & docid                 & 31  \\
& 2014-10-28 16:10:16 & view\_record          & 392 \\
& 2014-10-28 16:16:48 & search                & 10  \\
& 2014-10-28 16:16:48 & searchterm\_2         & 10  \\
& 2014-10-28 16:16:48 & resultlistids         & 10  \\
& 2014-10-28 16:16:58 & view\_record          & 9   \\
& 2014-10-28 16:17:07 & goto\_google\_scholar & 0  \\
\noalign{\smallskip}\hline
\end{tabular}
\end{table}

As specified in Section~\ref{sec:user behaviour analysis}, we are interested, in this paper, in specific stratagems namely \quotes{Citation Searching} (CS), \quotes{Footnote Chasing} (FC), \quotes{Keyword Searching} (KS), \quotes{Journal Run} (JR) and \quotes{Author Searching} (AS). These stratagems are present in our dataset such as \textit{view\_citation} aka CS, \textit{view\_references} aka FC, \textit{keyword\_search} aka KS, \textit{person\_search} aka AS and \textit{searchterm\_2} aka JR. They are distributed as described in Table~\ref{tab:StratDistribution} where most of the users performing them were not registered in the sowiport Digital Library. In fact, 99.30\% of the individual sessions in the SUSS 14-15 dataset and 99.37\% of the individual sessions in SUSS 16-17 were performed by non-registered users.

\begin{table}
\centering
\caption{Distribution of stratagems in both datasets}
\label{tab:StratDistribution}
\begin{tabular}{lp{2cm}p{2cm}p{2cm}p{2cm}}
\hline\noalign{\smallskip}
& \multicolumn{2}{l}{Registered Users} & \multicolumn{2}{l}{Individual Sessions} \\
\noalign{\smallskip}\hline\noalign{\smallskip}
Stratagem & SUSS 14-15 &  SUSS 16-17 & SUSS 14-15  & SUSS 16-17 \\
\noalign{\smallskip}\hline\noalign{\smallskip}
Citation Search   &   93     &  14   &  18,833  &  15,023 \\
Footnote Chasing  &   39	 &   11  &  1,520   &  673    \\
Keyword Search    &   161    &  40   &  27,126  &  9,547  \\
Author Search     &   181    &  43   &  50,957  &  17,312 \\
Journal Run     &   389   & 125   &  47,871  &   24,643 \\
\noalign{\smallskip}\hline
\end{tabular}
\end{table}

\subsection{Measurements}
\label{sec:measures}
To answer the first research question described in Section~\ref{sec:intro}, we analyse the sessions with the mentioned stratagems FC, CS, KS, AS and JR. 

For a session $S$ during which a set of interactions $\left \{ I \right \}$ is performed by the user, we define:
\begin{itemize}
\item $Strat$ is a stratagem such as FC, CS, KS, AS and JR
\item $Pos$ is a positive interaction\footnote{The following positive actions are considered as implicit relevance indicators.} present in our dataset among the following set
$\left \{ P \right \}$ described in \cite{Hienert:2016}:\\ {goto\_fulltext, goto\_google\_scholar, goto\_local\_availability, goto\_google\_books, view\_description, export\_cite, export\_bib, export\_mail, to\_favorites, \newline export\_search\_mail, save\_search, save\_search\_history, \newline save\_to\_multiple\_favorites}. 
\end{itemize}

In order to answer the second research question, we measure the precision of a stratagem before 
($P^{b}_{Strat}$) and after ($P^{a}_{Strat}$), so that we can verify if a certain stratagem has an influence on the conduct of a session. We verify if we can find more positive actions after using a stratagem compared against positive actions before its utilisation. 

\begin{equation}
P^{b}_{Strat}  = \left (\frac{\left | Pos \in \left \{ P \right \} \right |}{\left | I \right |}  \right )_b 
\label{eq:before}
\end{equation}

\begin{equation}
P^{a}_{Strat} = \left (\frac{\left | Pos \in \left \{ P \right \} \right |}{\left | I \right |}  \right )_a
\label{eq:after}
\end{equation}

To have an overview of a stratagem effect, we measure the \textit{Usefulness} ($U_{Strat}$) as the percentage of sessions containing positive actions among all the sessions in which the stratagems occur.  
This measure is inspired by the \textit{Global Usefulness} measure proposed by \cite{Hienert:2016}:
\begin{equation}
U_{Strat} = \frac{\left | s^{+}_{Strat} \right |}{\left | s_{Strat} \right |}
\label{eq:global}
\end{equation}
where $s^{+}(Strat)$ indicates session success in terms of positive actions occurrence after using a specific stratagem, and $\left | s_{Strat} \right |$ represents the number of sessions using a stratagem (\quotes{Footnote Chasing} or \quotes{Citation Searching}) no matter the type of user's interactions (positive or not).

\section{Results} 
\label{sec:results}
\begin{figure}[!h]
\begin{center}
	\includegraphics[width=1.0\textwidth, height=10cm]{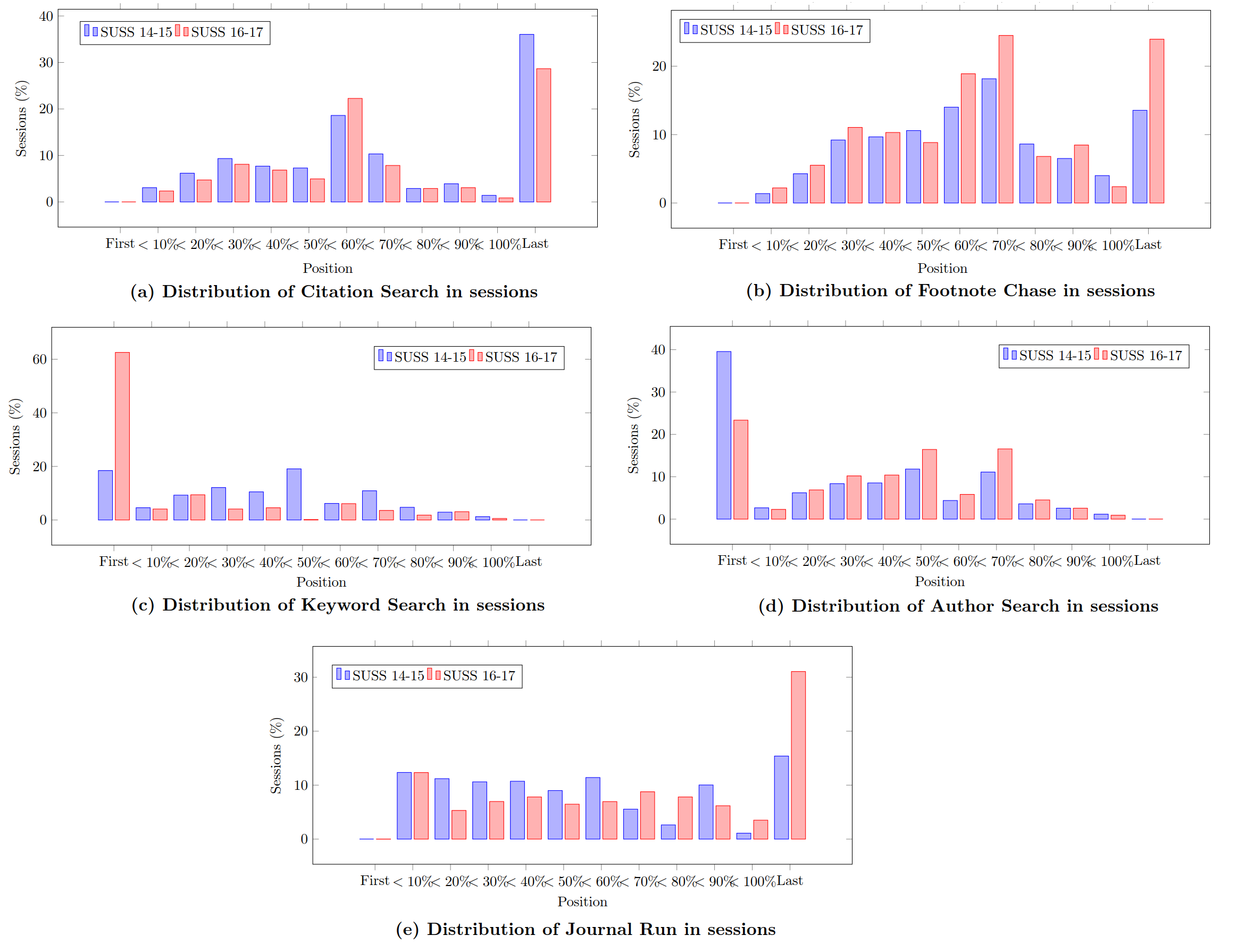}
	\caption{Number of sessions containing stratagems and the stage of their appearance in a session - (a), (b), (c), (d) and (e) for respectively Citation Searching, Footnote Chasing, Keyword Searching, Author Searching and Journal Run.}
	\label{fig:positions}
\end{center}
\end{figure}

Our results show the distribution of stratagems at different stages of the sessions (see Figure~\ref{fig:positions}). 
For both datasets, we observe that the stratagems are distributed similarly during a session. 
We note that the position of the stratagem differs from one session to another due to the difference of the sessions' length. We noticed from the user behaviour analysis that for \quotes{Citation Searching} (Figure ~\ref{fig:positions}-a) and \quotes{Footnote Chasing} (Figure ~\ref{fig:positions}-b), the stratagems appear mostly in the middle (18.61\% - 18.16\%) and the end of the sessions (28.66\% - 13.55\%) for the SUSS 14-15 dataset.

Besides, \quotes{Keyword Searching} (Figure ~\ref{fig:positions}-c) and \quotes{Author Searching} (Figure ~\ref{fig:positions}-d) appear mainly in the beginning of the sessions in both datasets. \quotes{Keyword Searching} and \quotes{Author Searching} are used at the beginning of a session with 18.45\% and 39.56\% for the SUSS 14-15 dataset and with 62.54\% and 23.36\% in the SUSS 16-17 dataset respectively.

While searching sowiport, the user typically checks the document content and metadata such as title, keywords, source. By clicking on a specific journal, the user expresses his/her interest in the source (the journal) and can then browse other documents from the same journal (compare results in \cite{KacemM18}). For this reason, \quotes{Journal Run} (Figure ~\ref{fig:positions}-e) was mostly used in the end of the sessions for both datasets SUSS 14-15 (15.39\%) and SUSS 16-17 (31.05\%). This result is in line with the findings in \cite{Carevic:2015} where the authors highlighted the need for an extended approach regarding \quotes{Journal Run}.

We observe no change in the distribution of the stratagems in the sessions using the second dataset SUSS 16-17 except for a slight change towards \quotes{Journal Run}. It was used mostly in the end of the sessions for both datasets but its usage in the middle of the sessions was enhanced more in the SUSS 14-15 dataset (11.41\%) than in SUSS 16-17 one (6.95\%).

We conclude that the usage of the stratagems was stable from the analysis of both datasets that cover two different periods. The difference in the occurrence stage of the stratagems could be explained by the fact that the users initially enter the system looking for keywords describing specific topics or searching for documents published by specific known authors they are interested in. After exploring results via indexed keywords, or authors, the users are more likely to investigate references, citations and sources (journals) of relevant documents before ending the retrieval sessions.
Current retrieval systems can take advantage of the difference in the moment of use of these stratagems by extending search facilities for those appeared in the end (especially FC, CS and JR). For instance, retrieval systems can provide a journal-based ranking/recommendation in addition to the document-based one. Moreover, the ranking of citations and references could be based on the similarity to the user's topic of interest or their popularity regarding the number of clicks. Lastly, citations or references that are more related to the current paper or query could be ranked in the top-N items.

In order to study the effectiveness of the stratagems \textit{Footnote Chasing}, \quotes{Citation Searching}, \quotes{Keyword Searching}, \quotes{Author Searching} and \quotes{Journal Run}, we measure their precision before and after their appearance during search sessions. This measure is based on a set of positive actions that are considered as relevance indicators \cite{Hienert:2016}.
In Table~\ref{tab:results}, we present the results of the measures described in equations~\ref{eq:before}, ~\ref{eq:after} and~\ref{eq:global}. To verify the impact of stratagems on the sessions' conduct in terms of positive actions (e.g. add to favorite, export citation, ...), we use as a baseline a set of 100 random sessions that do not include any of the stratagems (FC, CS, KS, AS or JR). We, thus, can check the presence of positive actions in the sessions when no stratagem is used, and if the presence effectively triggers more positive interactions within the search system. As no stratagem was used in the baseline sessions, we only measured the \textit{usefulness} which represents the percentage of sessions having relevance indicators (positive actions are described in detail in Section \ref{sec:measures}) compared to the total number of sessions (which is equal to 100\%).

\begin{table}[!htbp]
\centering
\caption{Evaluation of the effect of stratagem use on the sessions' search with significance using \textit{Student test} $p < 0.05$ for all compared pairs}
\label{tab:results}
\begin{tabular}{lllll}
\hline\noalign{\smallskip}
Stratagems &  $P^{b}_{Strat}$  & $P^{a}_{Strat}$ & Gain in Precision\tablefootnote{The gain is computed as the percentage of increase between the precision-after and the precision-before.} & $U_{Strat}$  \\
\noalign{\smallskip}\hline\noalign{\smallskip}
 SUSS 14-15 & & & & \\
 FC & 0.0284 & 0.1975 &  16.19\% &  77.24\% \\
 CS & 0.0197 & 0.1897 &  17\%    &  73.80\%\\
 JR & 0.0204 & 0.0459 &  2.55\%  &  41.57\% \\ 
 KS & 0.0321 & 0.0282 &  0.39\%  &  32.79\% \\
 AS & 0.0196 & 0.0289 &  0.94\%  &  27.36\% \\

 \noalign{\smallskip}\hline
 SUSS 16-17 & & & & \\
 FC   &  0.0236 &  0.1760  & 15.24\%   &  78.11\% \\
 CS   &  0.0409 &  0.2500  &  20.91\%  &  79.56\% \\
 JR   &  0.0395 &  0.0964  & 5.32\%    &  55.83\% \\
 KS   &  0.0353 &  0.0488  &  1.35\%   &  48.56\% \\
 AS   &  0.0246 &  0.0307  & 0.61\%    &  27.79\% \\

 \noalign{\smallskip}\hline
 Baseline\tablefootnote{The baseline is a set of 100 random sessions that do not include any of the stratagems.} & - & - & -  &  19\%\\
\noalign{\smallskip}\hline
\end{tabular}
\end{table}

From Table~\ref{tab:results}, we conclude that, for the dataset SUSS 14-15, all stratagems have a positive impact on the session performance in terms of positive actions appearance with different scales. This improvement is higher for FC and CS stratagems (16.19\% and 17\% respectively) compared to JR, KS or AS (2.55\%, 0.39\% and 0.94\% respectively). As for the global performance of all stratagems, the usefulness values of 77.24\%, 73.80\%, 41.57\%, 32.79\%, 27.36\% for respectively FC, CS, JR, KS and AS are considered as promising results as the value of this measure is in $\left [ 0,1 \right ]$ and the results are improved compared to the baseline with at least 8\%.

Considering the second dataset SUSS 16-17, we notice that the impact of the stratagems is similar to the dataset SUSS 14-15 in terms of usefulness and gain in precision. Sessions have different values of gain in precision with higher ones for FC and CS (3.95\% and 9.17\%) comparing to JR, KS or AS (5.32\%, 1.35\% and 0.61\% respectively). Regarding the usefulness values, we find that all of the stratagems have higher values than the baseline with also more significant values for CS and FC (78.11\% and 79.56\%) compared to JR, KS and AS (55.83\%, 48.56\% and 27.79\%).

In order to study the effect size, which is an approach to quantify the difference between two variables, we measured the \textit{Cohen's d} distance \cite{Cohen:1988}. This distance is used to identify how significant an effect is and to examine effects across variables. It is measured based on the difference between two means divided by a standard deviation for the data.
According to the results of Table \ref{tab:size effect}, we found large \textit{d} values indicating that the difference between the pairs of stratagems is large enough and consistent enough to be important for all pairs; except for those: KS-JR (d=0.37 for SUSS 14-15 and d=-0.18 for SUSS 16-17), AS-JR (d=0.21 for SUSS 14-15 and d=0.16 for SUSS 16-17), KS-AS (d=0.28 for SUSS 14-15 and d=0.38 for SUSS 16-17) and FC-CS (d=0.39 for SUSS 14-15).

\begin{table}
\centering
\caption{Effect size results using \textit{Cohen's d} measure using both datasets SUSS 14-15 and SUSS 16-17 where $^\ast$ indicates pairs with a small size effect}
\label{tab:size effect}
\begin{tabular}{llllll}
\noalign{\smallskip}\hline
 SUSS 14-15 & & & & &\\
 \noalign{\smallskip}\hline
& CS & FC & AS & KS & JR  \\
CS & - & - & - & - &  - \\
FC & \textbf{0.39}$^\ast$ & - & - & - & -  \\
AS & 0.9 & 0.51 & -  & - & -  \\
KS & 0.86 & 0.52 &  \textbf{0.28}$^\ast$ & - &-  \\
JR & -0.84 & -0.49 &  \textbf{0.21}$^\ast$ & \textbf{0.37}$^\ast$ &-  \\ 

\noalign{\smallskip}\hline
 SUSS 16-17 & & & & &\\
\noalign{\smallskip}\hline
  & CS & FC & AS & KS & JR  \\
CS & - & - & - & - &  - \\
FC & 0.94 & - & - & - & -  \\
AS & 1.54 & 0.63 & -  & - & -  \\
KS & 1.30 & -0.57 &  \textbf{0.38}$^\ast$ & - &-  \\
JR & -0.92 & -0.64 &  \textbf{0.16}$^\ast$ & \textbf{-0.18}$^\ast$ &-  \\ 
\noalign{\smallskip}\hline
\end{tabular}
\end{table}

We see for both datasets that in most of the cases the amount of positive actions which appeared before a stratagem utilisation are lower compared to positive actions which appeared after the stratagem utilisation. We can present an improvement in terms of occurrences of positive actions when a stratagem is employed by the users and thus conclude that these stratagems lead to more successful sessions and positive interactions with the system.

In Table~\ref{tab:posneg}, we present the different ways in which a stratagem affects the conduct of a session by measuring the difference $Diff(a,b)$ between the precision before ($P^{b}_{Strat}$) and the precision after ($P^{a}_{Strat}$) their use. The effect could be positive ($Diff(a,b) > 0$), non-positive ($Diff(a,b) < 0$) or neutral ($Diff(a,b) = 0$).
We can see that both stratagems CS and FC affect the sessions mostly in a positive way. This means that the use of a stratagem influences the user behaviour and makes him/her more interactive with the system in a beneficial way. In spite of the fact that the stratagems AS, JR and KS have a lower positive effect in both datasets, we can see that they have a neutral effect compared to the non-positive one. In fact, the absence of positive actions does not mean a negative conduct of a session because the user is always interacting with the system using moves and tactics that are not judged as positive relevance indicators but not specified as negative either.

\begin{table}
\centering
\caption{Impact of the stratagems (FC, CS, KS, AS and JR) on sessions}
\label{tab:posneg}
\begin{tabular}{p{2cm}lllll}
\hline\noalign{\smallskip}
 & FC & CS & KS & AS & JR \\
\noalign{\smallskip}\hline\noalign{\smallskip}
& \multicolumn{3}{l}{SUSS 14-15} \\
\noalign{\smallskip}\hline\noalign{\smallskip}
Positive Effect ($Diff>0$) & 67.83\% & 68.97\%  & 19.82\% & 18.38\% & 33.2\%  \\
Non-positive Effect ($Diff<0$)& 9.41\% & 4.83\% & 12.96\% & 8.98\% & 8.56\%\\
Neutral Effect ($Diff=0$)& 22.67\% & 26.20\% & 67.22 \% & 72.64\% & 58.4\% \\
\noalign{\smallskip}\hline\noalign{\smallskip}
& \multicolumn{3}{l}{SUSS 16-17} \\
\noalign{\smallskip}\hline\noalign{\smallskip}
Positive Effect ($Diff>0$) & 66.23\% & 34.48\%  & 21.84\% & 17.67\% & 27.42\% \\
Non-positive Effect ($Diff<0$)& 11.87\% & 20.44\% & 14.84\% & 10.12\% & 17.76\% \\
Neutral Effect ($Diff=0$)& 21.89\% & 35.08\% & 63.32\% & 72.21\% & 54.82\%  \\
\noalign{\smallskip}\hline
\end{tabular}
\end{table}

\begin{figure}
\begin{center}
	\includegraphics[scale=0.5]{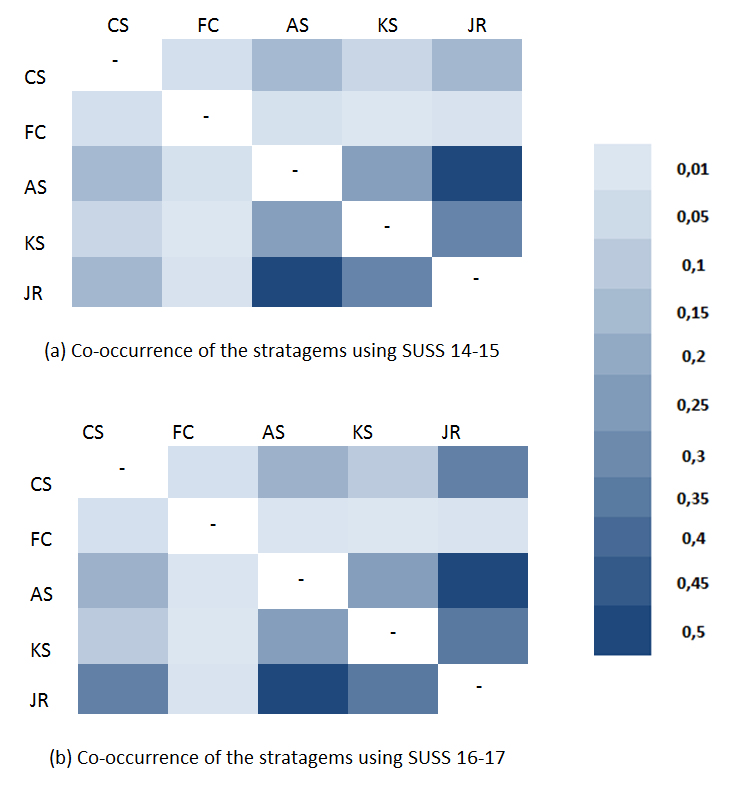} 
	\caption{Co-occurrence matrix between the stratagems CS, FC, KS, AS and JR for both datasets}
	\label{fig:matrix}
\end{center}
\end{figure}

In order to analyse the correlation between the stratagems, we present in Figure ~\ref{fig:matrix} the matrix of co-occurrences for both datasets. We notice that the highest co-occurrence is between JR and AS followed by the co-occurrence of JR and KS. However, a weak correlation between FC and both AS and KS is noted. 
For the SUSS 16-17 dataset, the percentage of co-occurrence is different as the size of the dataset is not the same. We notice that the correlation JR-AS is the highest for both datasets. The correlation between stratagems has increased in SUSS 16-17 compared to SUSS 14-15 for all of them except for FC-CS and FC-AS where it has decreased by 1.46\% and 29.43\% respectively. Specifically, the correlation JR-CS has significantly increased by 110.8\% in SUSS 16-17 compared to SUSS 14-15.

We conclude that the \quotes{Journal Run} is the stratagem presenting the highest correlations with other stratagems. In fact, users who are looking for a specific source (journal) are also looking for the authors who published these documents and the topics presented in their descriptions. 

\section{Discussion}
Understanding the user behaviour in Digital Libraries is essential to create useful and effective systems. We investigated the user behaviour in an academic search engine from the perspective of five types of stratagems \quotes{Footnote Chasing} (FC), \quotes{Citation Search} (CS), \quotes{Keyword Search} (KS), \quotes{Author Search} (AS) and \quotes{Journal Run} (JR). We considered the mentioned stratagems in two datasets \textit{SUSS 14-15} and \textit{SUSS 16-17} that were collected over a period of a couple of months each. The first dataset contains 484,449 while the second one contains 208,556 retrieval sessions. 
These stratagems were first introduced by \cite{Bates} whereupon many studies explored their usefulness in the context of Digital Libraries and academic search. We find it interesting to observe the user behaviour towards these stratagems as they are essential for the search session conduct and, thus, academic search system providers can take advantage of them and include them in the ranking process and the interface.
In fact, most of the existing systems in different fields use mainly the personalisation/recommendation based on the content similarity between the query and the research paper and do not provide a contextualised ranking based on the paper's metadata such as citations, references, etc.
Through this log analysis, we suggest considering further elements to provide adapted search to the user and to improve his/her experience within the system. Moreover, the findings of this log analysis can provide insights on how to improve the Digital Library's interface \cite{mahoui:2001}. In fact, a stratagem-based improvement could be achieved in the search interface. For instance, the interface could include a citation-based query interface in addition to the document-based one. Also, the recommended items, currently based on research paper similarity, could be established through citations or references. Lastly, a research paper could include many citations or references that are not directly related to the paper's topic. Thus, a DL can display the top-N citations or references for which the content match best the query terms.

As an answer to the first research question \textit{Which usage patterns can be observed from specific search stratagems or moves?}, we analysed the frequency and stage of usage of these stratagems in the search sessions. We noticed that the stratagems appear within different scales. Taking both datasets into account, AS and JR are the most used stratagems for both datasets, compared to CS, FC and KS. 
Thus, users are more likely to search for specific authors and their material or the source of specific papers (journals) rather than looking for indexed keywords, references or citations. Specifically, Footnote Chasing was the less frequently used stratagem in both datasets. In fact, FC and CS have not the same chance to get used compared to AS, KS or JR due to the non-availability of references/citations for some documents in the underlying collection.
As for the position of the stratagems, we found that the stratagems are used differently and appear in different stages in a search session. Considering both datasets SUSS 14-15 and SUSS 16-17, the stratagems AS and KS appear mostly in the first positions of the sessions, while CS, FC and JR appear mostly in the last positions.

Regarding the second research question \textit{How successful are the retrieval sessions which contain specific stratagems?}, we used the two measures \textit{precision} and \textit{usefulness} in order to verify the influence of stratagems on the session. We found that all the sessions have been useful to enhance the users' interactions within both sowiport datasets ($U_{Strat} > 0$ and $P_{Strat} > 0$). In the light of the stratagem effect that we measured by the difference of the precision after and before a stratagem use, we conclude that, for both datasets, CS and FC have a positive effect while KS, AS and JR have a neutral effect in most of the sessions.

Regarding the last research question \textit{How significant are the changes in the usage of stratagems using the same retrieval system in two different periods?}, we found that FC, KS and AS have decreased in terms of frequency by 4.23\%, 23.30\% and 25.92\% while CS and JR have increased by 73.42\% and 11.98\% respectively (see Table \ref{tab:StratDistribution}). We noticed that AS has the highest frequency of usage followed by JR in both datasets compared to all actions performed by the users. However, in terms of the user behaviour towards the stratagems and the influence of their utilisation on the sessions' conduct, we have seen that they are similar for both datasets with an important value of usefulness for all stratagems, and a higher positive impact using both of the stratagems FC and CS compared to KS, AS, and JR (see Tables \ref{tab:results} and \ref{tab:posneg}). As for the co-occurrence, we conclude that the user behaviour is also stable as JR-AS is the most co-occurred pair compared to the other pairs of stratagems in both datasets while the pair JR-CS has increased in SUSS 16-17 compared to SUSS 14-15 in terms of co-occurrence (see Figure \ref{fig:matrix}).

We conducted this log analysis in order to answer the research questions addressed and because the findings can help the DLs support the user during his/her search. In fact, the system support could be achieved at different levels. Among the actions proposed by DLs, the five stratagems studied could be enhanced in a way to improve the user experience and to adapt the results to his/her interest. Personalised approaches could be provided based on the user's previous actions and the stratagems he/she performed (see an example of contextual browsing for DL in \cite{carevic2018}). The improvement could be achieved, for instance, by presenting a better ranking of items for the user.
In this direction, a citation-based ranking/recommendation can provide a user with the top-N most related citations to his query, or a journal-based ranking if he/she is checking a specific edition of a journal \cite{Carevic:2015}. 
Another improvement could be achieved by granting the user access to citations and references (as they are widely used in DLs). As the recommendation is needed in advanced search systems to ensure a rapid, easy and perceptive navigation, the DLs can provide intuitively recommended items through an optimised visualisation. 
For instance, systems can provide a recommendation of papers related to the user's topic of interest implicitly without an extra effort from the user.

\section{Conclusion}
In this paper, we investigated the use of five search stratagems in the sowiport search engine. In fact, studying the user behaviour towards stratagems can enhance the user-system interactions and lead to more useful academic search engines \cite{Carevic:2015}. 
Using the SUSS datasets, we examined the frequency and stage of use of such stratagems as well as their impact on sessions. We verified whether their utilisation can lead to successful sessions. We measured the success of sessions based on the difference between the precision before and after using a stratagem. Both precisions are obtained thanks to the number of positive relevance indicators compared to all actions in a session \cite{Hienert:2016}. 
In addition, we compared the results obtained from the measures and analysed the stability of the user behaviour towards the utilization of stratagems on two different datasets.

In conclusion, we realised a log analysis and an observation of the user behaviour towards five stratagems \quotes{Citation Search}, \quotes{Footnote Chase}, \quotes{Author Search}, \quotes{Keyword Search} and \quotes{Journal Run}. The aim of this observation is to provide a better understanding of the stratagems and to highlight the need of information systems to consider the user's context in order to adapt the results to their clicked keywords, authors searched, references, citations or journals explored. In fact, these stratagems can be explicitly provided in the user interface by, for instance, recommending papers of a specific journal (clicked by the user) and related to the topic of interest. A journal-based ranking could be also achieved as a search facility worthy to be considered by search engines \cite{Carevic:2015}. In addition, the references and citations could be enhanced in a way to provide statistics on citing and cited papers as well as click-able links so that the user can easily check them without typing a new query. Currently, systems solely offer the list of references and citations. However, this ranking could be optimized by enhancing the positions of the papers that are more related to the current document to be put in the top ranks.

As future work, we need to go beyond log analysis and perform user studies in order to compare user feedback with the findings of this study. In addition, it will be beneficial to include personalised items based on the stratagems performed by the user in the search system's interface \cite{carevic2018}. This allows comparing the usage and the impact of stratagem-based ranking compared to the standard ranking.

\begin{acknowledgements}
This work was funded by Deutsche Forschungsgemeinschaft (DFG), grant no. MA 3964/5-1; the AMUR project at GESIS together with the working group of Norbert Fuhr. 
The AMUR project aims at improving the support of interactive retrieval sessions following two major goals: improving user guidance and system tuning.
We thank Julia Achenbach for her proof reading of the final version of this paper.
\end{acknowledgements}

\bibliographystyle{spmpsci}      
\bibliography{stratagems.bib}   

\end{document}